\documentclass[preprint,preprintnumbers,amsmath,amssymb]{revtex4}

\usepackage{graphicx}
\usepackage{dcolumn}
\usepackage{bm}

\begin{document}

\title{ Transparency effect in the emergence of monopolies in social networks}
\author{ A. H. Shirazi $^1$, A. Namaki $^{2}$, A. A. Roohi  $^{3}$, G. R. Jafari $^{4,1}$ \thanks{Email: g_jafari@sbu.ac.ir}$^\dag$ ,
 \\
 {\small $^1$ Computational Physical Sciences Research Laboratory, School of
Nano-Science, Institute for Research in Fundamental Sciences (IPM),
Tehran, Iran} \\
{\small $^2$ Department of Financial Management, Faculty of
Management, University of Tehran, Tehran, Iran} \\
{\small $^3$ Engineering Physics Department, Ecole Polytechnique,
Montreal, H3C 3A7, Canada }  \\
{\small $^4$ Department of Physics, Shahid Beheshti University,
G.C., Evin, Tehran 19839, Iran} }
\date{\today}

\begin{abstract}
Power law degree distribution was shown in many complex networks.
However, in most real systems, deviation from power-law behavior is
observed in social and economical networks and emergence of giant
hubs is obvious in real network structures far from the tail of
power law. We propose a model based on the information transparency
(transparency means how much the information is obvious to others).
This model can explain power structure in societies with
non-transparency in information delivery. The emergence of ultra
powerful nodes is explained as a direct result of censorship. Based
on these assumptions, we define four distinct transparency regions:
perfect non-transparent, low transparent, perfect transparent and
exaggerated regions. We observe the emergence of some ultra powerful
(very high degree) nodes in low transparent networks, in accordance
with the economical and social systems. We show that the low
transparent networks are more vulnerable to attacks and the
controllability of low transparent networks is harder than the
others. Also, the ultra powerful nodes in the low transparent
networks have a smaller mean length and higher clustering
coefficients than the other regions.

\end{abstract}
\maketitle

\section{Introduction}

Power law degree distribution of different kinds of complex
networks, were issued by many researchers in recent years
\cite{3,4,newman,namaki,Amir}. Barabasi-Albert model was a basic
attempt to describe this phenomena. The main concern of this model
was the preferential attachment. The model showed that preferential
attachment in a growing network leads to a power law degree
distribution, as well as a random attachment that leads to an
exponential degree distribution. In recent years, there were many
different variations of this model \cite{3,4}. For describing the
behavior of real systems, the main focus of these models is to
reproduce the growth process in real networks. In essence, they
describe the dynamical mechanisms that produce the network. The
Dorogovtsev-Mendes-Samukian $(DMS)$ model is a complete form of BA
model that premises the presence of the initial number of nodes
\cite{6}. Krapivsky et.al introduced a model with a nonlinear
preferential attachment probability \cite{7}. Klemm-Eguiluz (KE)
proposed a model known as structured scale-free model that describes
the dynamic growth of the networks based on the memory of the nodes
\cite{8}.

Despite these models, there are other methods for describing the
growth process of real networks \cite{9,10,11,12,13,14,15}.
However, in some social structures, we observe that power (an
interpretation of the effects on total network) condensates in some
nodes which breaks the scale free behavior \cite{a1, a2}. The
deviation from scale free behavior can be explained by applying some
modifications to the BA model \cite{a3, a4, a5}. Such deviations
were shown in Sornette's works on power law's distributions which he
called them "dragon kings" \cite{16,17}.

In sociology, there is a phenomenon called the Matthew effect which
describes the behavior of those nodes who have power. In economy the
dragon nodes are called economical power whilst in the society is called
political power. The Matthew effect is the phenomenon where "the rich get
richer and the poor get poorer" \cite{18,19,20}. In the networks,
power can be realized by the nodes' degree, betweenness or closeness
\cite{power}. In the BA model, everyone has full information about the other
nodes, so information is available for them to attach to
nodes with high degree. However, in social networks this kind of
information, diffuses through the network itself. The information
diffusion, like all other diffusions, can be subjected to some
restrictions. These restrictions will cause uncomplete and
non-accurate information. The rate of this diffusion can affect the
structure of our network as the system grows. In the case of social
networks, we found out that this rate has a crucial role that
directly reflects in the structure of power in a society. Also, in
economic networks, this diffusion rate has a close connection for
describing the competitiveness of the economic environment.

In this paper, we propose a definition of information transparency
for nodes degree distribution. Then, we make the modified
preferential attachments based upon this definition. The properties
of these networks based on their different diffusion
rates, are also studied.

\section{Modified Model}

The assumption of the preferential attachment in BA model is based
on adding a new vertex which attaches to vertex \emph{i} with a
probability $\prod$ that depends on the degree $k_{i}$, so:

\begin{equation}
\prod(k_{i}) = \frac{k_{i}}{\sum_{i}k_{i}}.
\end{equation}
The BA model assumes the availability of the nodes degree
information for each new node introduced to the system. However, in
the modified model, we consider that the new vertex first connects
to node $i$ randomly without any prior information about the
degree of that node. Then, it finds out about other nodes' degree
through the node $i$. Since the degree known by node $i$
has passed through several edges, it does not express the exact
degree of the other nodes. This is because the information about the
nodes' degree changes "$r$" times, each time if passes an edge. We call
this an information diffusion. Hence, we introduce the term $k_j^i$
which is the degree of node $j$ viewed by $i$ (the node $j$'s degree
has been diffused through the network before reaching node $i$).
$k_j^i$ is $k_j$ that has diffused $d$ times:
\begin{equation}
k_j^i = k_jr^{d_{ij}},
\end{equation}
which $d_{ij}$ is the shortest path length between nodes $i$ and $j$
and the new node connects according to connectivity $k_j^i$
(Usually, most reliable information obtained from a node, are
information which have come from the shortest path). So the
probability that a new node which is connected randomly to node $i$
can make a connection to node $j$ as:
\begin{equation}
\prod(k_j^i)=\frac{k_j^i}{\sum_{j}k_j^i}.
\end{equation}
Then each new node makes $m$ new edges to remain in the network.
With the aid of this model, various deviations from scale free
behavior can be explained by different values of diffusion rate. It
is obvious that, \emph{r}=$0$ will result in a randomly growing
graph and \emph{r}=$1$ represents the BA model. All above steps could be
summerize as follows:

\emph{\\
 1) Start with a small core network. (In our simulation we
start with a $m$-clique which $m$ is the edge number that connects
the new node to the network.) \\
2) Choose a random node $i$. \\
3) Calculate the $k_{ij} = k_j r^{d_{ij}}$, which is the node $j$
degree's viewed by node $i$. \\
4) Connect a new node with $m$ edges to the other nodes, with
preferential attachment according to $k_{ij}$. \\
5) Refresh $k$s and $d$s. \\
6) Return to step (2). \\}

A schematic example is presented to clarify this model. A new person
in town does not have accurate information about important
(well-known) people of the city. He may come to a person randomly and ask
him about the others. His judgment about the others is crucially
dependent on how accurateness of the information he had gathered from
the people he had met. If we have a perfect transparency in information, i.e.
($r = 1$), then accurate information to make
connections throughout the network is available. This network growth follows
BA model where the degree distribution posses a power-law behavior. However, with perfect
non-transparency in networks' information delivery, i.e. $r = 0$, he
has no useful information about anyone and connects randomly to
another node in the network, which means random growth and
exponential behavior in nodes' degree distribution. Our results show
that, between these two limits, there are rates that the networks
with these rates have nodes with amazingly high degree which is
interpreted as emergence of ultra powerful hubs in social networks.
We consider edges between nodes to be homogenous which means they
are all as of the same kind with the same diffusion rates.

\section{Results \& discussion}

The main purpose of this paper is to study the effect of information
distortion in the construction of networks. Based on the above
model, we can construct different networks in respect to different
diffusion rates ("$r$") where $r$ is the parameter that makes this
distortion in information delivery. If $r$ changes from 1 to 0, it
makes the nodes' degree to show lesser than the actual nodes degree.
On the other hand, if $r$ changes from 1 to higher values, it causes
exaggeration and overestimation of the nodes degree.

In Fig.(1), we have developed networks based on this modified model
for some different diffusion rates. It is obvious that for $r=0$ the
network is a random graph and for $r=1$ it is the same as the BA model.
In $r=0.05$ some powerful hubs emerge and in $r=5$ observe
random behavior again.

In Fig.(2) we have depicted the degree distribution for different
diffusion rates, based on the assumption of growing the network by
adding one by one nodes with $m=5$ number of new edges, which are
added for each new node. We have added nodes till $N = 10^{4}$ (Fig.2
a, d, b, e)and for showing that the finite size effect is not an important matter
in our growing process, we have shown results for $N = 10^{5}$ (Fig.2 c, f).

It is observed by increasing the diffusion rates form zero to one,
the degree distribution moves from a random exponential network to a
Barabasi-Albert power law  model. Between these two points, there
are some diffusion rates in which the networks with these rates have
nodes with a high degree that cause deviation in the networks degree
distributions from the power law behavior.
$r=0$ is the state of no transparency in information delivery.
This is equivalent to the state of random growing network with
exponential degree distribution.

$r=1$ is the state with a complete and accurate transfer of
information throughout the network which reproduces the BA model
with power law degree distribution.

In the low diffusion rates, lower than $0.01$, we still observe
exponential behavior for the majority of nodes with lower degree.
Even though we have random behavior in the rest of the network with
maximum degree about $10^{2}$, the emergence of nodes with amazingly
high degree about $10^{4}$ is a noticeable fact, as the rate goes
above $0.005$.
As the diffusion rate increases to $0.05$, the network starts to
show a power-law behavior while the nodes with ultra high degrees are
still present. As it is obvious, there is a power law behavior in
the beginning of some distributions (by eliminating the powerful
nodes from the distribution), where the slope is $4.2±0.03$
in $r = 0.05$, and by $r$ increasing to $1.5$, it decreases to $3$.
Either every where $r>1.5$ the distributions do not have the power-law behavior.

If we continue to increase the diffusion rate, the model gets more
and more closer to the BA model. After that, we studied the networks
with $r>1$. If we continue to increase the diffusion rate above $1$
as shown in Fig.2 (b, e), it can be seen that the powerful hub is
disappearing as the rate goes up, and the whole system shifts towards an
exponential behavior in a connectivity structure. In other words,
the system shifts to the random growth as the diffusion rate
increases above $1$. In social interpretation, the network has
experienced exaggeration of information which results in a random
behavior. So, this shows that the fake information in systems is
equivalent with no information.

In essence, there are four distinct parts for the proposed model:
non-transparent, low transparent, perfect transparent and
exaggerated regions.

By considering different information diffusion rates, different
social, economical and political situations involved in
information delivery of societies can be modeled. The emergence
of high degree nodes is interpreted as the emergence of powerful
hubs (high power nodes which are dominant in size and importance) in
social networks as a result of low-transparency in information
delivery.

Information is not only a tool for being dominant, but is the power
itself. In some cases, information sources, adjust the diffusion
rate on purpose in favor of a party. These are societies with the
power, condensed in these monopoles as they try to maintain the
power with the aid of censorship or supportive actions from
government. Some famous examples of these structures are
undemocratic governments, where power condensates in the hands of a
powerful political elite group. Also, because of the low-transparent
competitive environment, sometimes firms emerge as central nodes in
economic networks. With total transparency in information delivery,
which leads to societies without monopolies, it can be considered as
an ideal model. This is the perfect case and most of the times, real
phenomena are deviated from this ideal model.

As the diffusion rate increases to values above one, the new in town
is in a situation of information overflow, which will lead him to
the same result of having no or less accurate information. In other
words, having no information is the same as having huge amount of
information which is not accurate or is exaggerated. In real cases,
societies are sometimes bombard with propaganda which can totally
restructure power systems to other random structures, the case that
happens in some government structures. Some firms such as medias, which
controll the amount of informal statements in societies, can
restructure the power system to their desirable shape by controlling
the amount or accuracy of the information.

In order to have a better sense about the model, we have plotted
Fig.(3a) which shows the maximum degree of developed networks by the
model for different diffusion rates and the average of this maximum
degree perceived by the other nodes. This perception shows what is
the opinion of the other nodes about the maximum degree in the
network on average or what is their estimation of the size of the
powerful hub. The figure is in logarithmic scale, and the vertical
error bars show the diversity of opinion about the size of this hub.

\subsection{General network properties}

 The emergence of nodes with high degree, will
decrease the mean length, and increase the clustering coefficients
of low degree nodes connected to them  as shown in Fig.(3b,c).
The decrease in the mean length, as a result of emergence of ultra
powerful hubs, is explained as a consequence of connection of most
nodes to one or more nodes with ultra high degree which act like
bridges in the network (Fig.(3b)). In economic and social networks,
the mean length can show the speed of diffusion of crises among
nodes of the system. Where this item is very small, it shows that
the crisis can diffuse very fast in the network. As we can see in
low transparent region, the mean length is much smaller than the
other regions and we expect that the diffusion of any event among
nodes has high speed. The increase in the average of the clustering
coefficient of the nodes, even in nodes with low degree, is the
result of the connectivity to powerful hubs as a dense core. The
mean clustering coefficient of the economic and social networks,
shows the remaining probability of the crisis in different groups of
clusters for a long time. In the low transparent region, it is
obvious that the mean clustering coefficient is much larger than the
other regions in Fig.(3c).

\subsection{Network robustness}

Here comes a peculiar question that whether a controlled censorship
really works in keeping the powerful hubs from falling down. If
these ultra powerful hubs encounter failure for any reason, will
the network face a serious break down? To answer this question, we made
attacks on the networks with different diffusion rates. Attack means
removing nodes from the network due to some defined rules. There are
several types of attacks \cite{attack}. We attack network nodes on
 their degree ranks. We eliminate nodes from the top degree to the
bottom, and after each step, evaluate the giant component size and
the mean size of the other components in the network \cite{attack}.
Results have been depicted in Fig.(3d) and Fig.(3e) for networks
with different diffusion rates. As we can see, the giant component
size and the average size of the other components are more sensitive
to the attacks in the low transparent networks. It is well known
that a random network is more robust than a scale free one against
targeted attacks \cite{attack}. But, the main finding is that the
low transparent networks ($0<r<1$), are more fragile, in comparison to
 both random and scale-free networks.

 The giant component
size decreases rapidly in the low transparent networks, compared to
other networks. In other words, in scale free models, the
society is less dependent to a special person or node. But in low
transparent systems, networks are highly dependent on special nodes
which are the center of connectivity.

In economic networks, there are some powerful hubs that are
considered to be "too big to fail" \cite{taleb,stern,wesel}. In
essence, emergence of these hubs is the result of low transparent
competitive environments. These financial institutions are so large
and so interconnected that their failure will be harmful to the
economy. This concept results in the belief that these firms should
become recipients of beneficiary financial policies from governments
or central banks to keep them alive. It is thought that these firms
have high-risk and are able to leverage these risks based on
the supportive actions. This term has emerged as an important
concept since $2007 - 2010$ in global financial crisis, that
bankruptcy of some giant companies has systemic effects on the total
economy.

\subsection{Controllability}

In real world, if we want to control a system, one method is to
control the set of driver nodes, which driving them by an external
signal results to control the systems' dynamic. Liu et. al used a
method, that was named maximum matching, for finding the minimum
number of driver nodes to attain full control of a complex network
with a dynamic behavior in its nodes \cite{liu}. In this paper, we
assume that our proposed networks are directed from the previous to
the latest nodes. Fig.(3f) shows that the number of the driver nodes
that must be controlled in low transparent region is much more than
the other regions, and in essence this makes more cost for
controlling the network.

In low transparent networks, the social capital comes from the
powerful hubs and the other nodes do not have common perceptions
about each other. So, controlling these networks forces more cost to
the powerful hubs. In low transparent societies, sometimes social
cohesion has different structures based on relationships between
powerful hubs and the other nodes in the social network. Most of the
cohesion is because of the existence of the hubs. So in these
networks, structural cohesion (the minimum number of members who, if
removed from a group, would collapse the group \cite{m}) is smaller
than the other networks.

\section{Conclusion}

In this paper we have presented a new method for generating social
networks. In this modified model, we have emphasized on the
diffusion rates as a mean for measuring the information transparency
in social and economical systems. The main interesting features of
the model is symmetry breaking of nodes degree due to both
exponential and power law distributions, despite of homogeneous
primary conditions. This model shows the emergence of different
groups of networks based on the different types of diffusion rates.
This view can model the reality of the social and economical
systems. In these systems, there are ultra powerful hubs that leads
to deviation from power law behavior and scale free concept. We have
computed the mean length and the clustering coefficients of
the networks based on different diffusion rates. It can be seen that
there are indirect relations between the diffusion rates and the
mean lengths, but there are direct relations between the clustering
coefficients and the diffusion rates. Also, we investigated the
behavior of the networks' structures with respect to the attack on
the powerful hubs, and was seen that the networks with low
diffusion rates are more sensitive to the attacks. Then, we
investigated the controllability of the networks. Our results showed
that the networks in low transparency region have more driver nodes
and are harder to control than the other regions.

\begin{acknowledgments}
The authors would like to thank Shahin Rouhani for his very helpful
comments and discussions, and Sara Zohoor and Soheil Vasheghani Farahani for helping to edit the
manuscript.
\end{acknowledgments}

\begin{figure*}[t]
 a) \includegraphics[width=7cm,height=6cm,angle=0]{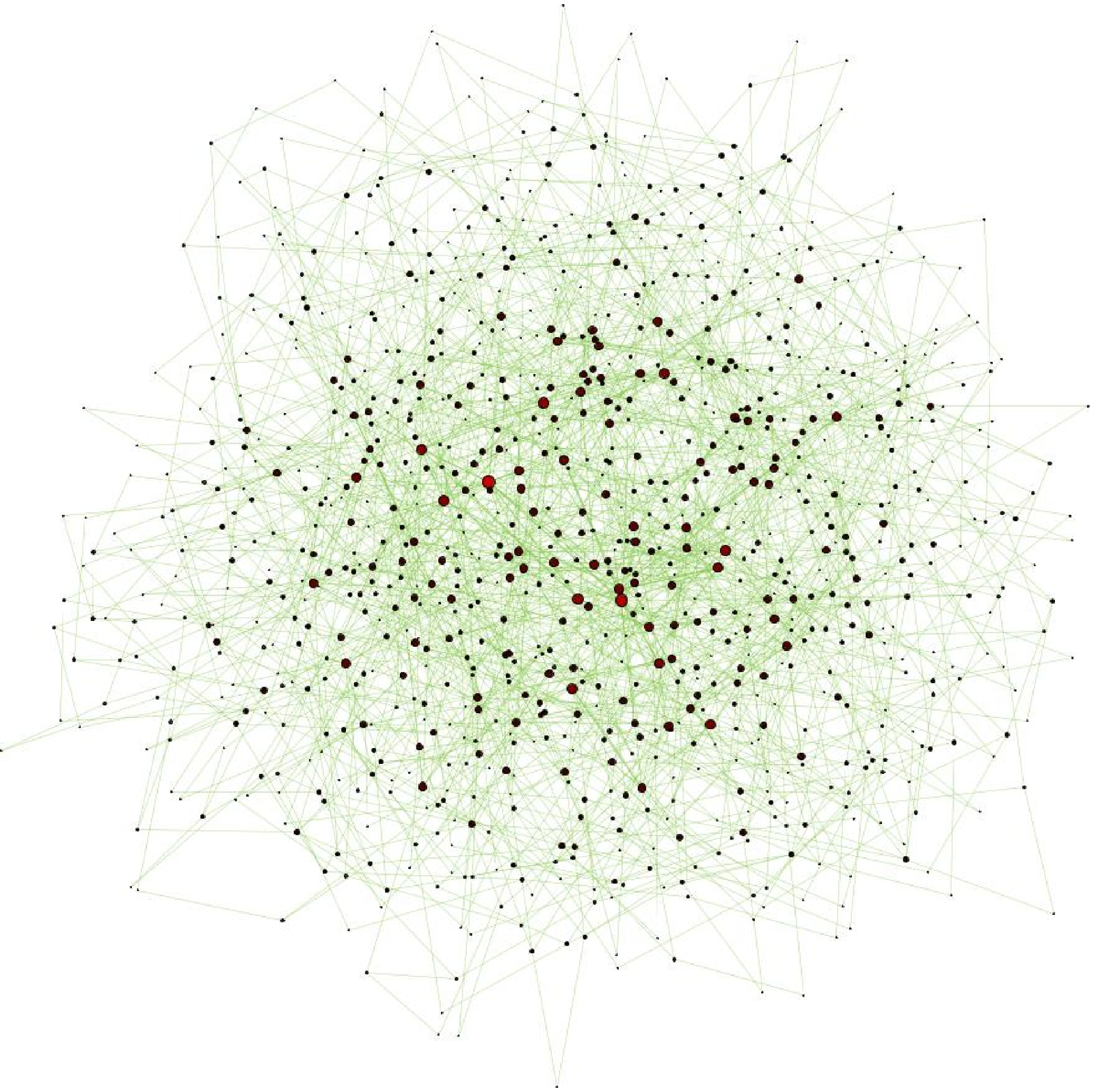}
 b)\includegraphics[width=7cm,height=6cm,angle=0]{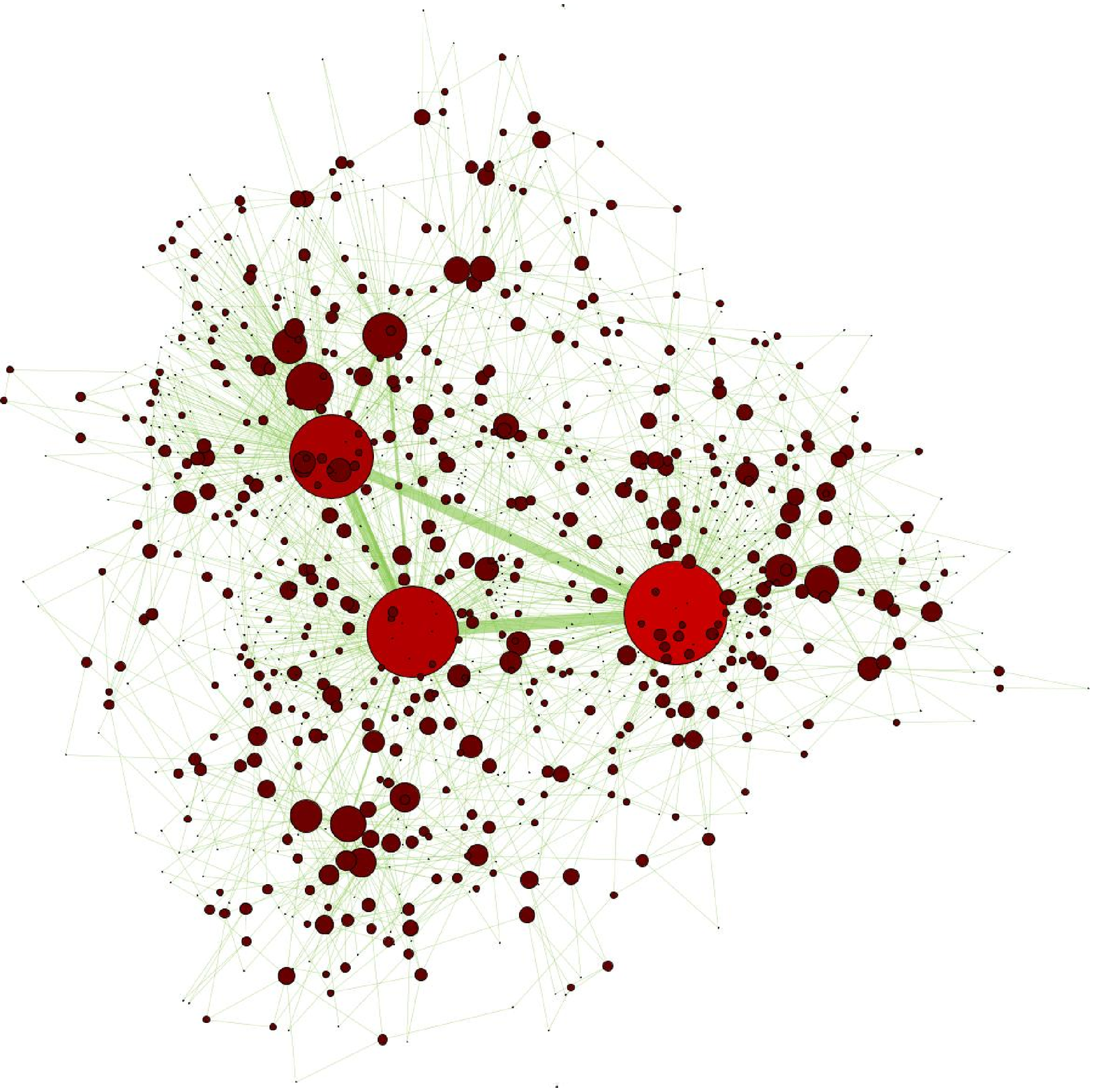}
\\
 c) \includegraphics[width=7cm,height=6cm,angle=0]{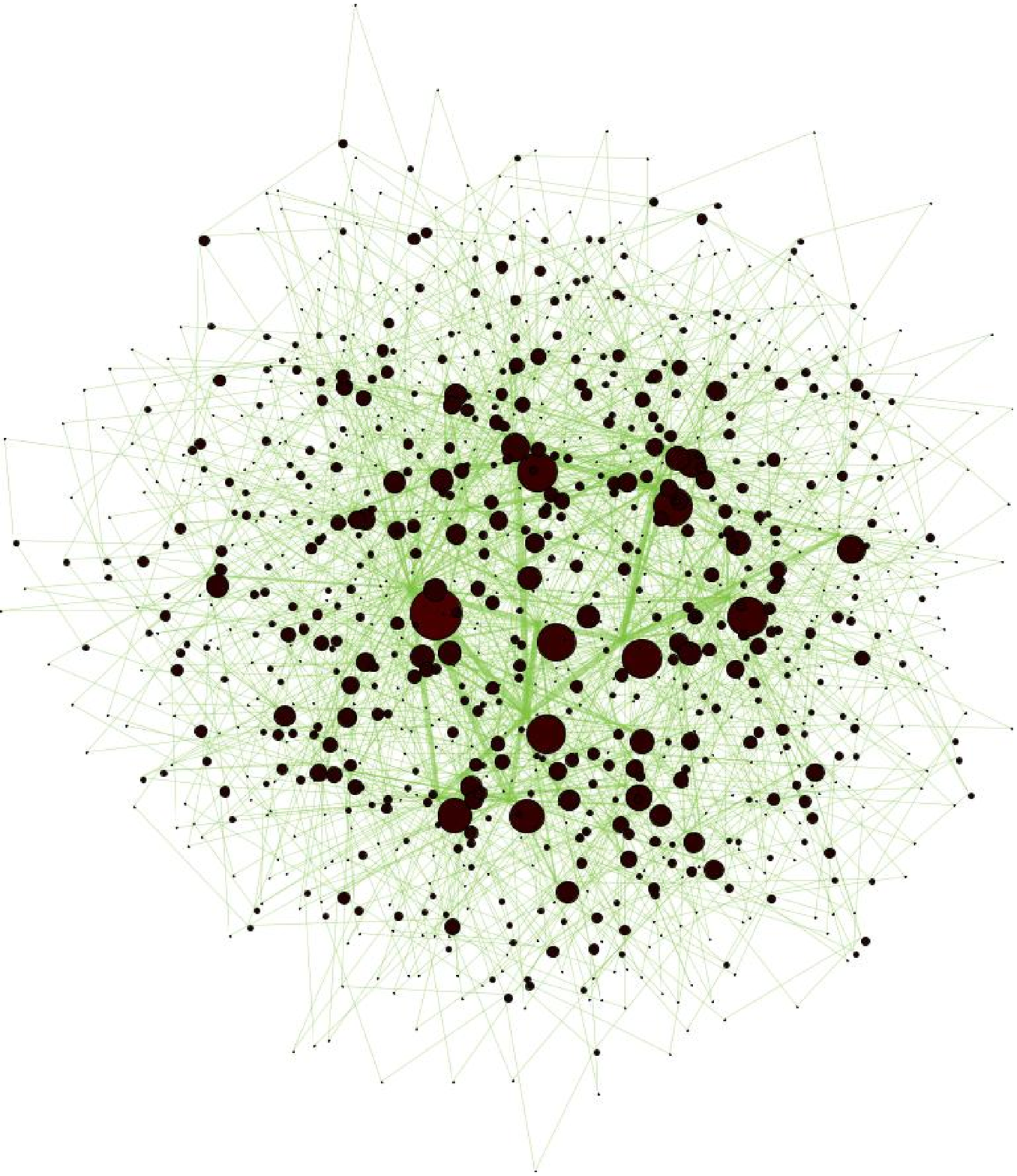}
 d) \includegraphics[width=7cm,height=6cm,angle=0]{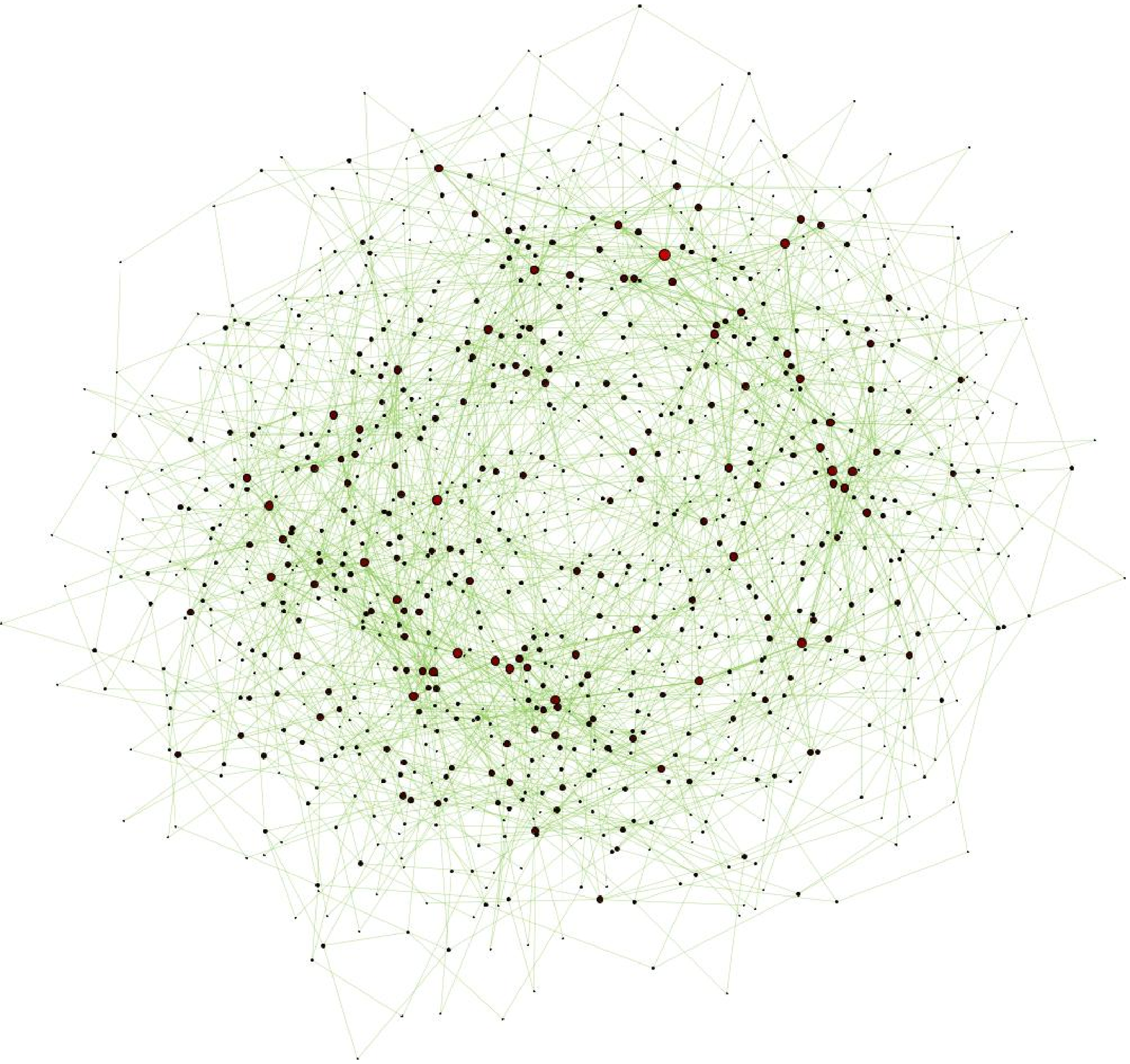}
\caption{Network samples for different diffusion rates \emph{r}, a)
0, b) 0.05 c) 1, d) 5. The diameter of nodes show their
degrees, which created by \cite{netfig}.}\label{fig1}
\end{figure*}
\begin{figure*}[t]
a)\includegraphics[width=5cm,height=12cm,angle=0]{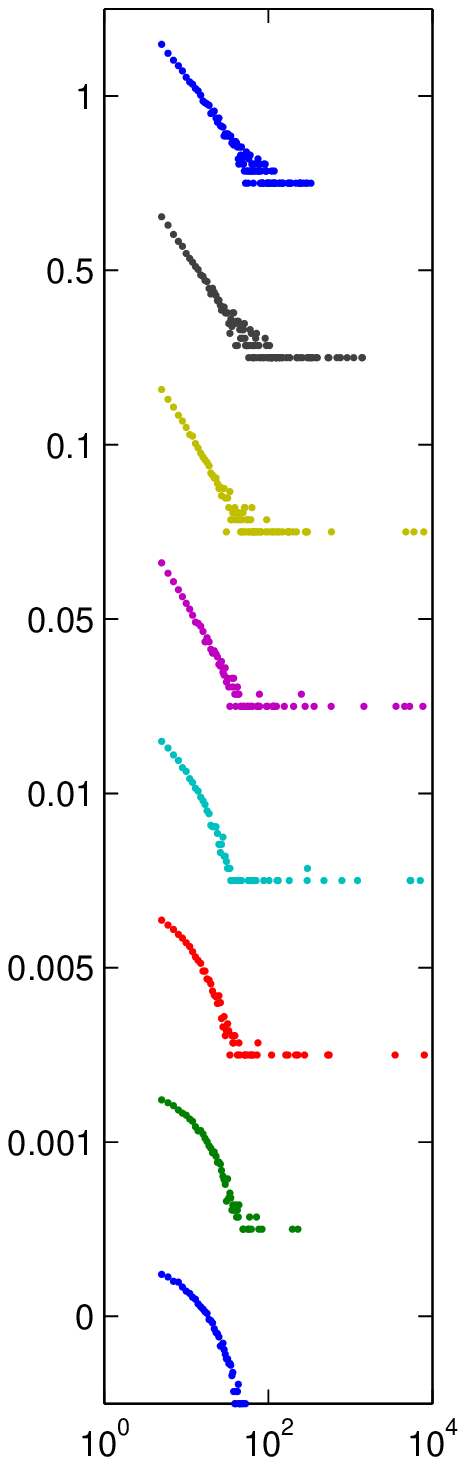}
b)\includegraphics[width=5cm,height=12cm,angle=0]{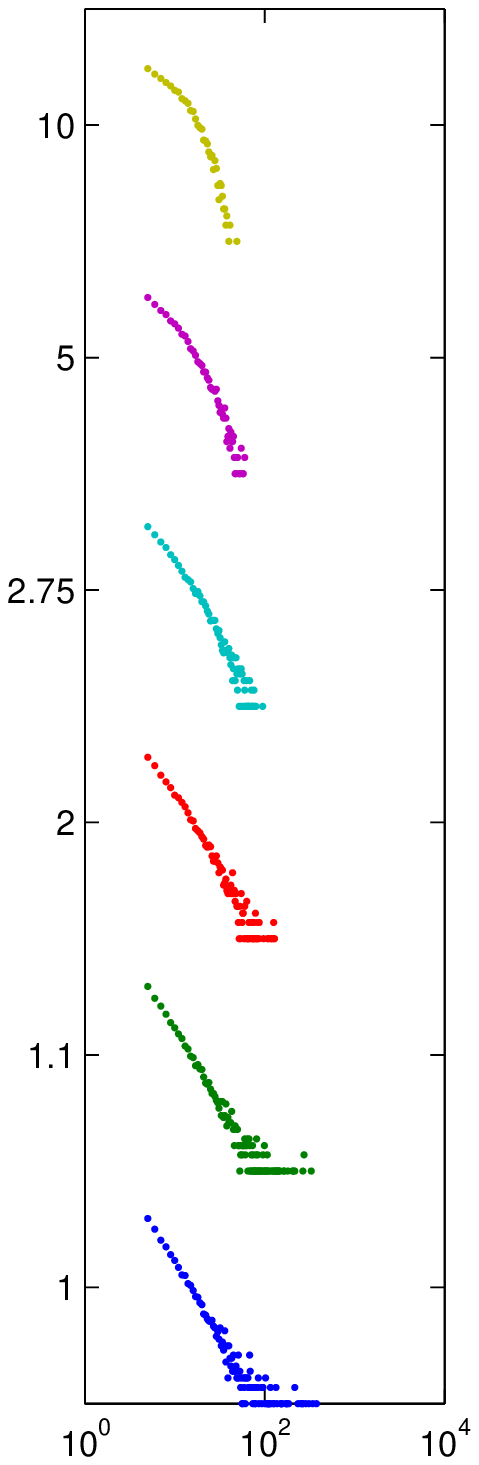}
c)\includegraphics[width=5cm,height=12cm,angle=0]{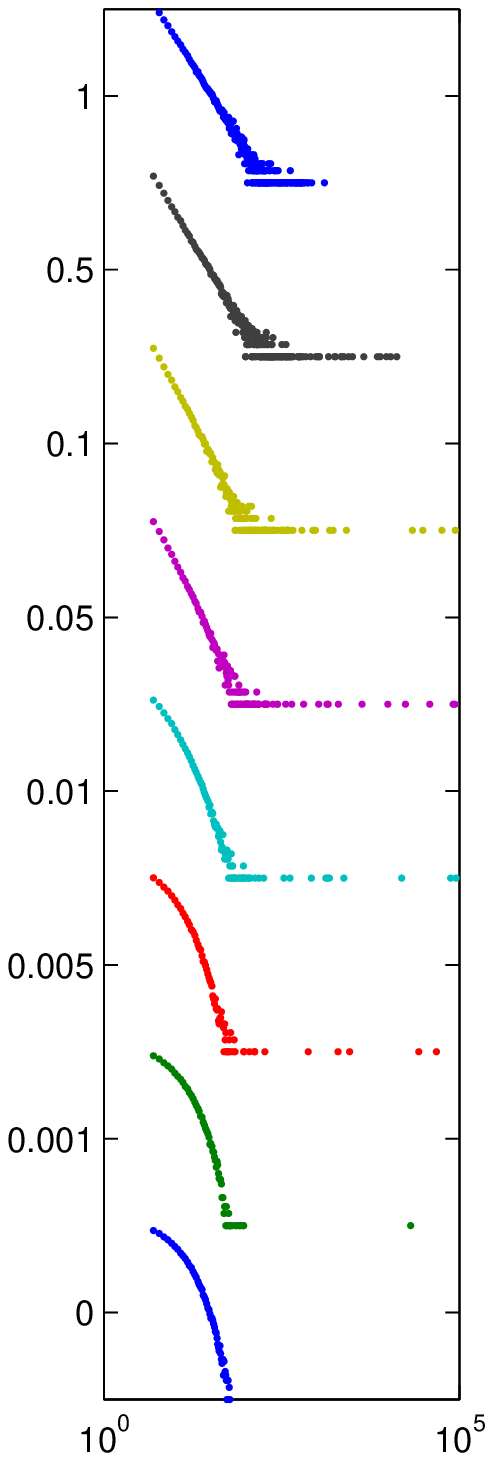}
\\
d)\includegraphics[width=5cm,height=4cm,angle=0]{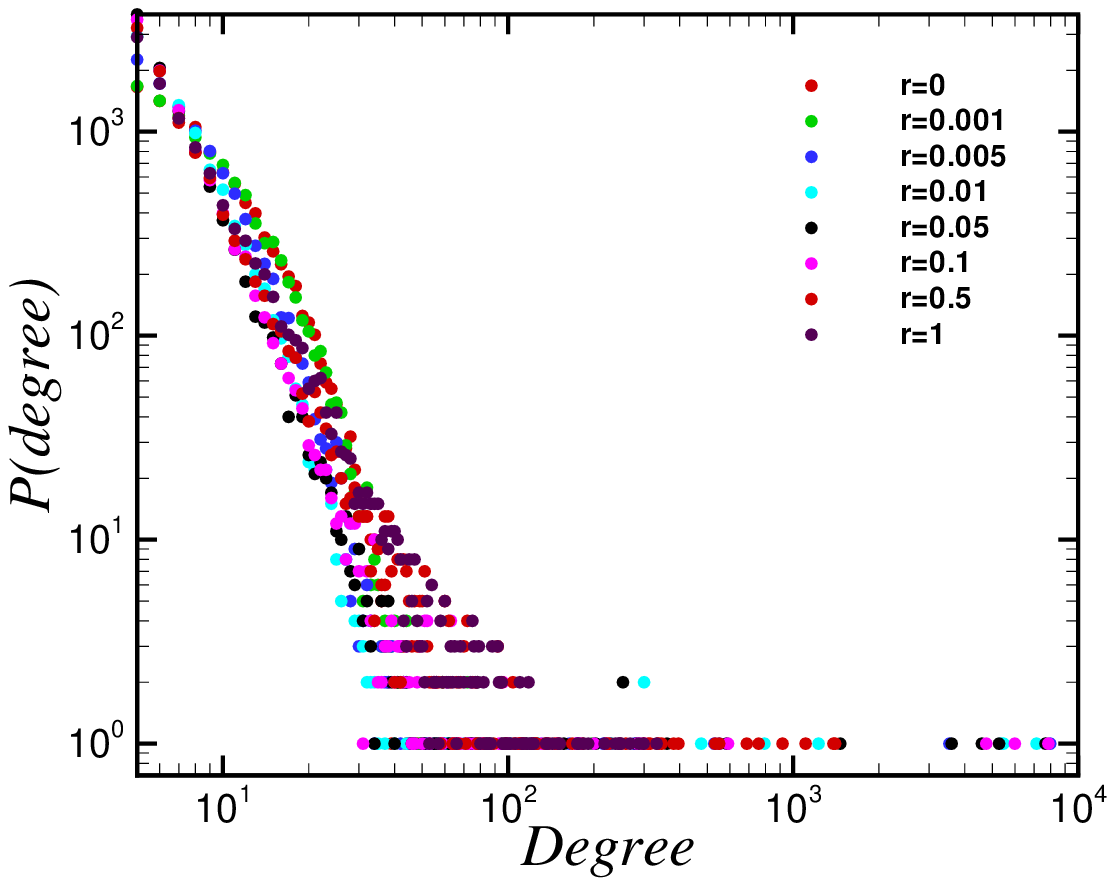}
e)\includegraphics[width=5cm,height=4cm,angle=0]{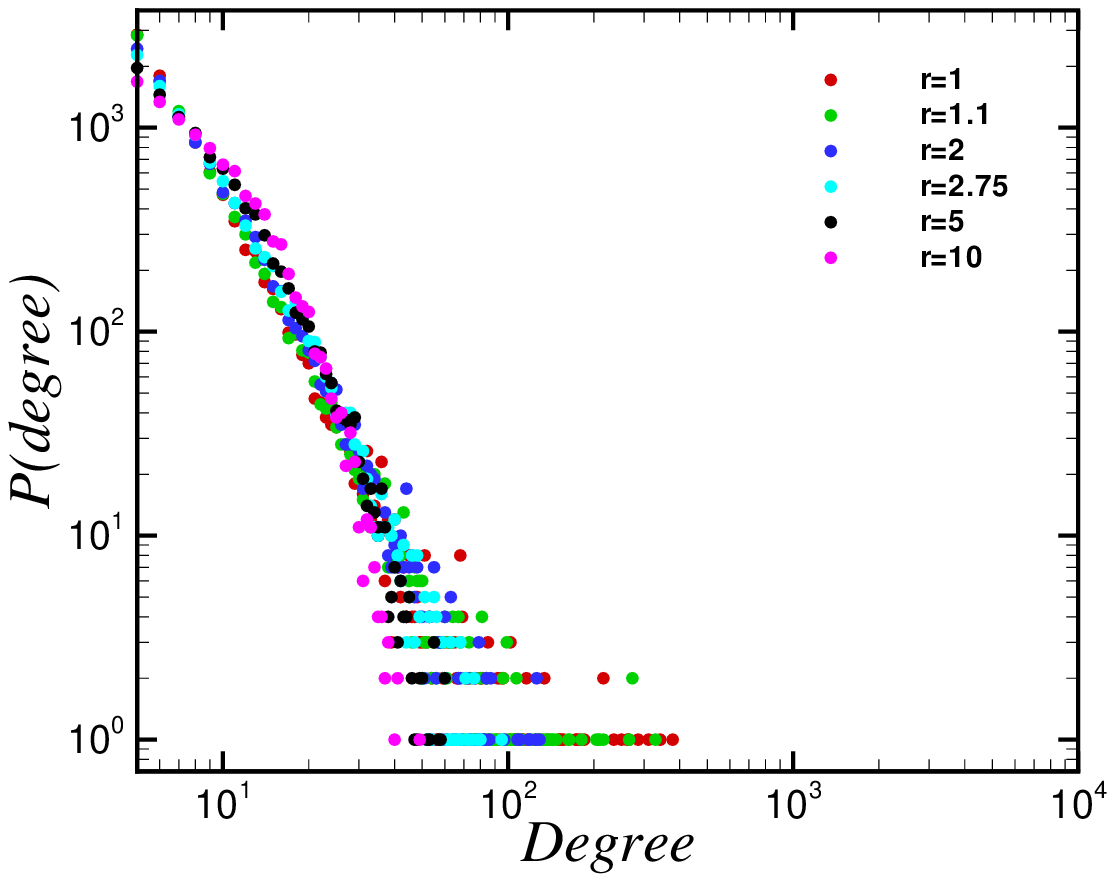}
f)\includegraphics[width=5cm,height=4cm,angle=0]{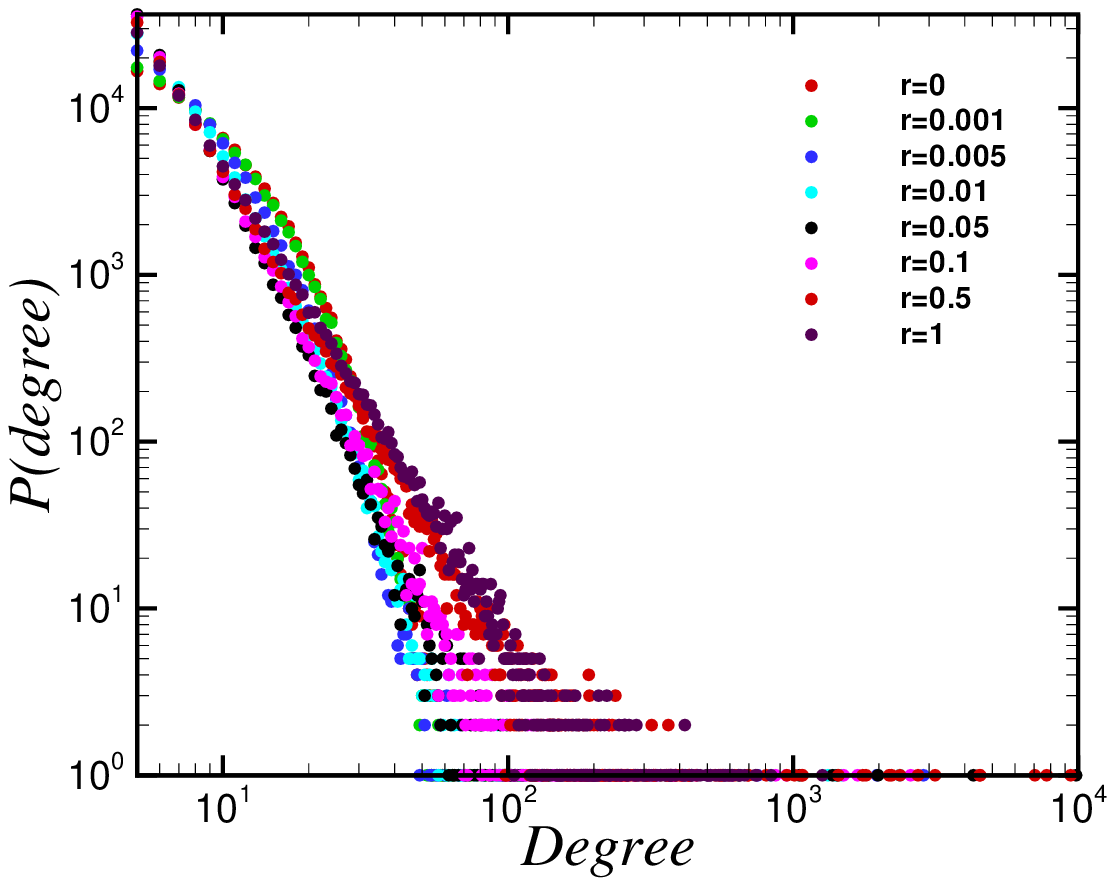}
\caption{The degree distribution for different values of $\emph{r} =
{(0, 0.001, 0.005, 0.01, 0.05, 0.1, 0.5, 1)}$ for the number of
nodes equal to (a)10000 and (c) 100000. (b) The degree distribution
for different values of $\emph{r} = {(1, 1.1, 2, 2.75, 5, 10)}$.
For more illustration of units and sclaes, the figures (d,e,f) is added
to show degree distributions in one plane.}\label{fig2}
\end{figure*}
\begin{figure*}[t]
a) \includegraphics[width=6cm,height=4.5cm,angle=0]{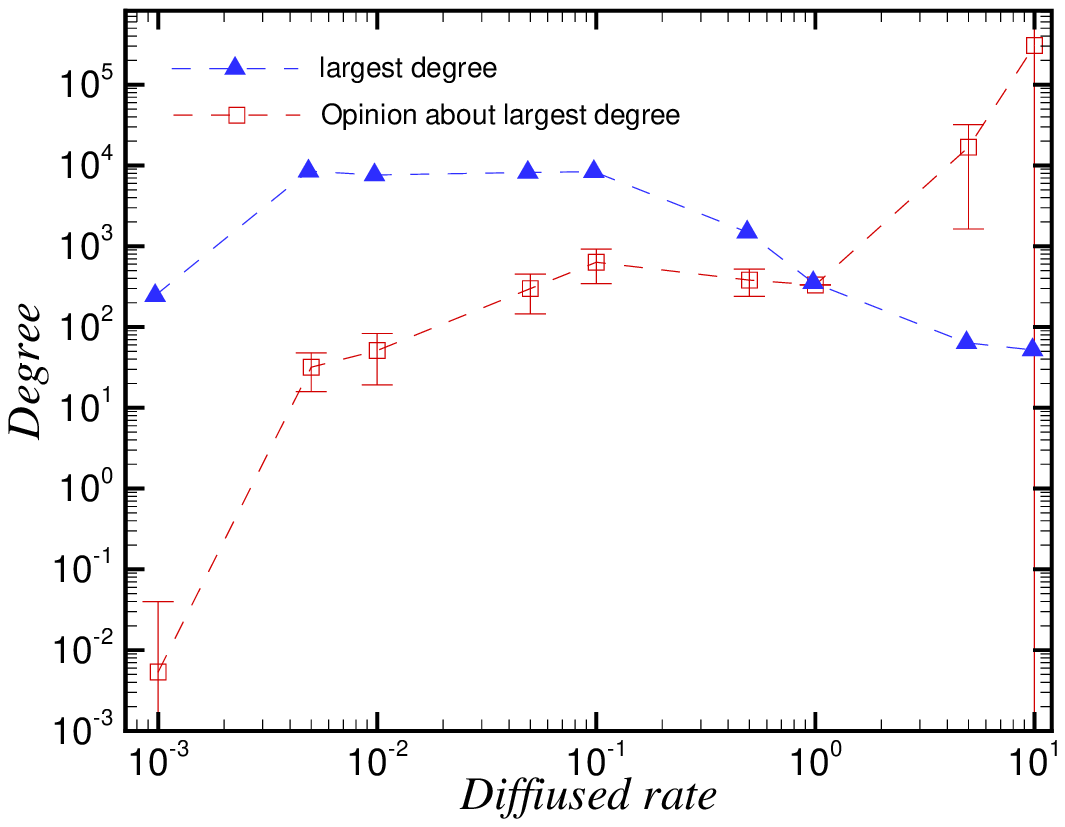}
b)\includegraphics[width=6cm,height=4.5cm,angle=0]{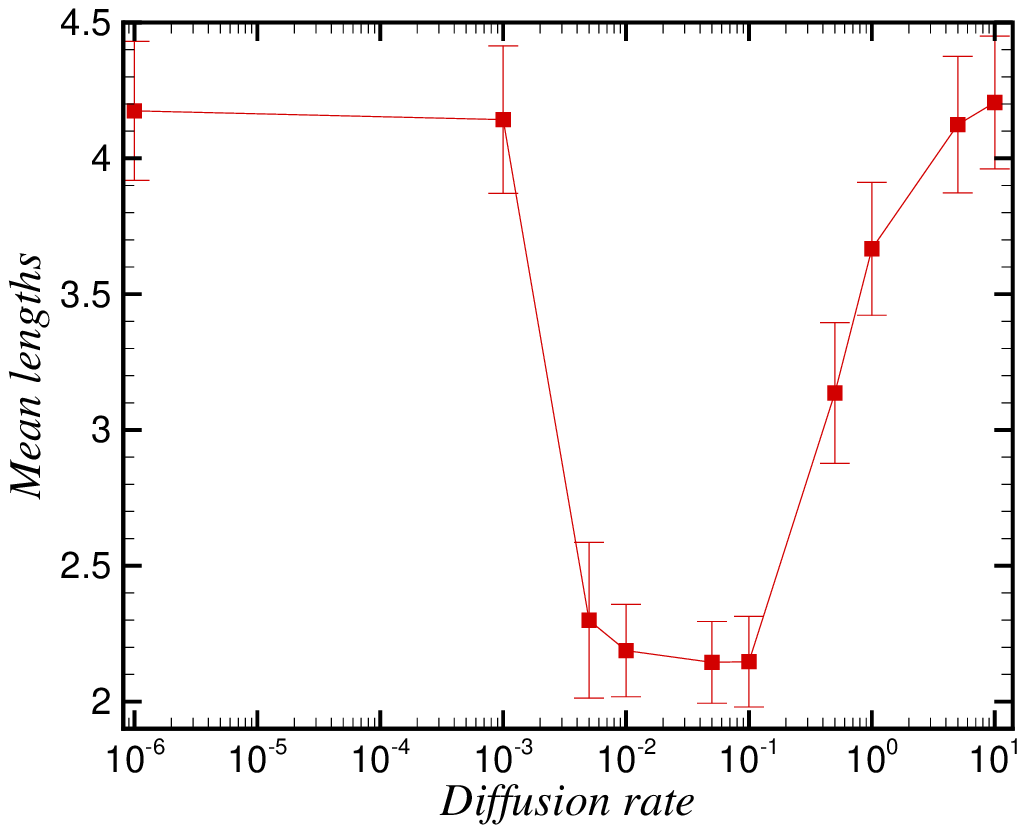}
c)\includegraphics[width=6cm,height=4.5cm,angle=0]{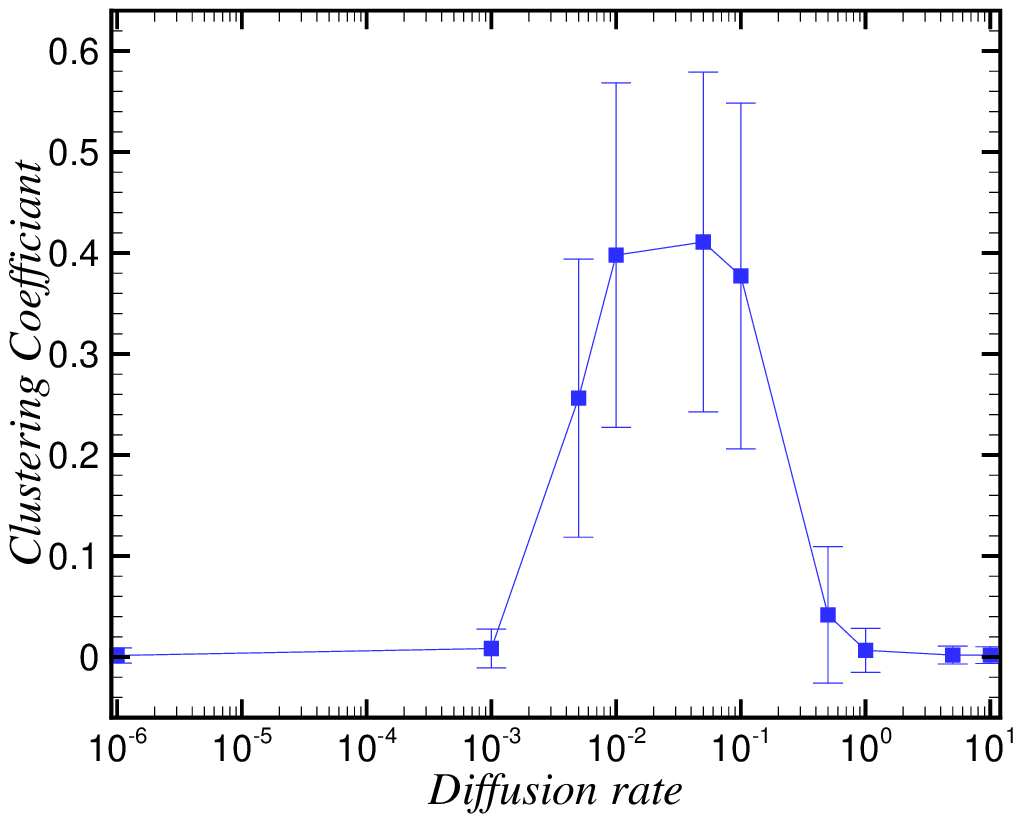}
d)\includegraphics[width=6cm,height=4.5cm,angle=0]{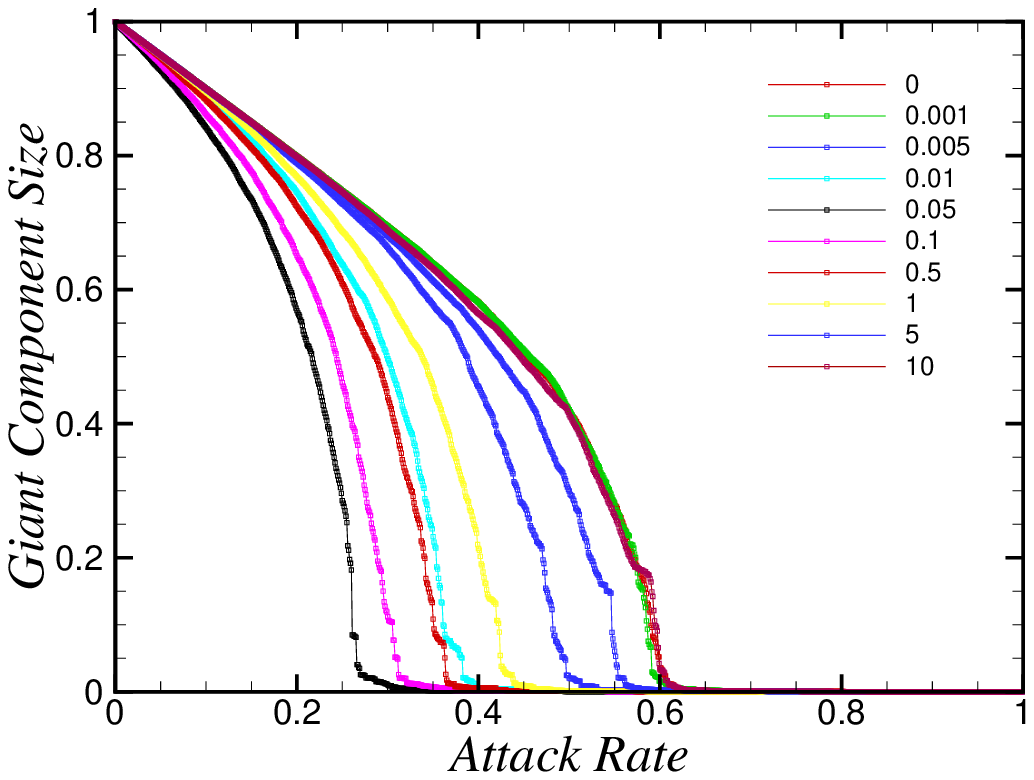}
e)\includegraphics[width=6cm,height=4.5cm,angle=0]{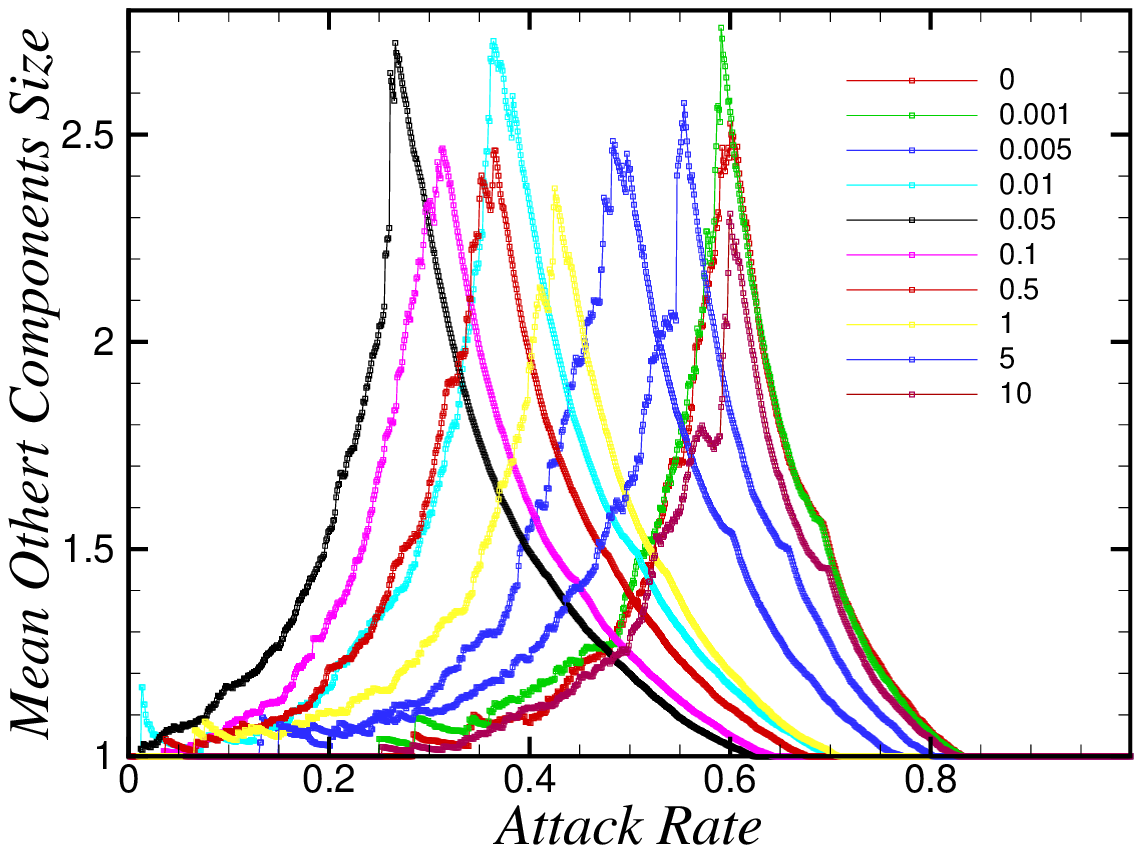}
f)\includegraphics[width=6cm,height=4.5cm,angle=0]{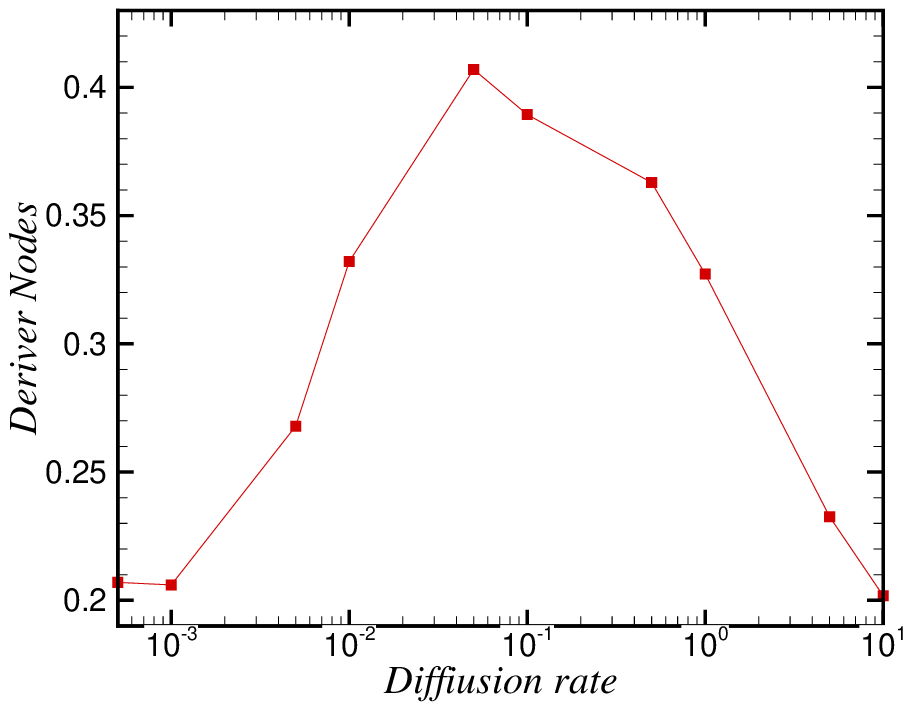}
\caption{a) Largest node degree in different diffusion rates and the
average of the other nodes' opinion about their degrees. b) The mean
of the shortest paths in the networks with different diffusion
rates. c) The mean clustering coefficients in the networks with
different diffusion rates. d) The giant component size vs removed
nodes' percentage. Nodes are removed due to the rank of their
degrees. Networks with low transparency show a more sensitivity to
the attacks e) The mean of the other component sizes vs the removed
nodes' percentage. f) Driver nodes (nodes which should be controlled
externally) in different diffusion rates.}\label{fig3}
\end{figure*}

\end{document}